\documentclass{aa}
\usepackage{graphicx}

\begin{document}
   \title{Emission of SN~1006 produced by accelerated cosmic rays}

   \subtitle{}

   \author{E.G.Berezhko
          \inst{1}
          \and
          L.T.Ksenofontov
          \inst{1}
          \and         
          H.J.V\"olk
          \inst{2}}

   \offprints{H.J.V\"olk}

   \institute{Institute of Cosmophysical Research and Aeronomy,
                     31 Lenin Ave., 677891 Yakutsk, Russia\\
              \email{berezhko@ikfia.ysn.ru}
              \email{ksenofon@ikfia.ysn.ru}
         \and
             Max Planck Institut f\"ur Kernphysik,
                Postfach 103980, D-69029 Heidelberg, Germany\\
             \email{Heinrich.Voelk@mpi-hd.mpg.de}
             }

   \date{Received month day, year; accepted month day, year}

     \abstract { The nonlinear kinetic model of cosmic ray (CR)
acceleration in supernova remnants (SNRs) is used to describe the
properties of the remnant of SN~1006. It is shown, that the theory fits
the existing data in a satisfactory way within a set of parameters which
is consistent with the idea that SN~1006 is a typical source of Galactic
CR nucleons, although not necessarily of CR electrons. The adjusted
parameters are those that are not very well determined by present theory
or not directly amenable to astronomical observations. The calculated
expansion law and the radio-, X-ray and $\gamma$-ray emissions produced by
the accelerated CRs in SN~1006 agree quite well with the observations. A
rather large interior magnetic field $B_\mathrm{d}\approx 100$~$\mu$G is
required to give a good fit for the radio and X-ray synchrotron emission.
In the predicted TeV $\gamma$-ray flux from SN~1006, the $\pi^0$-decay
$\gamma$-rays, generated by the nuclear CR component, dominate over the
inverse Compton (IC) $\gamma$-rays, generated by the CR electrons in the
cosmic microwave background. The predicted source morphology in high
energy $\gamma$-rays roughly corresponds to that of the synchrotron
emission. The predicted integral $\gamma$-ray flux $F_{\gamma}\propto
\epsilon_{\gamma}^{-1}$ extends up to energies $\sim 100$~TeV if CR
diffusion is as strong as the Bohm limit. Only if the interior magnetic
field is much lower in the SNR, $B_\mathrm{d}\approx 10$~$\mu$G, then the
observed $\gamma$-ray emission can be due to the accelerated electron
component alone. In this case, not plausible physically in our view, the
lowest permissible value of the electron to proton ratio is rather high,
and the maximum individual energy and total energy content of accelerated
nucleons so small, that SN~1006 can not be considered as a typical source
of the nuclear Galactic CRs.
    }
   
   \authorrunning{Berezhko et al.}
   \titlerunning{Emission of SN 1006}
   \maketitle
%

\section{Introduction}

In the last years significant efforts have been made to obtain direct
observational evidence whether the Galactic cosmic rays (CRs) --
relativistic nucleons and electrons -- are indeed generated in supernova
remnants (SNRs). The expected $\pi^0$-decay $\gamma$-ray emission,
produced in nearby SNRs by the accelerated protons in their collisions
with thermal gas nuclei, is marginally high enough to be detectable by the
present generation of imaging atmospheric Cherenkov telescopes (e.g. Drury
et al. \cite{dav94}; Naito \& Takahara \cite{naitot}; Berezhko \& V\"olk
\cite{bv97}). Positive results of such observations would constitute a
necessary condition for a dominant role of SNRs in the production of the
Galactic CRs and of their energy spectrum up to the knee energy $\sim
10^{15}$~eV.

If SNRs accelerate nucleons at least to ~100 TeV/n as it is required for
Galactic CRs (e.g Berezhko \& Ksenofontov \cite{berk}), then in smooth
extension of the expected hadronically generated $\pi^0$-decay
$\gamma$-ray peak at 67.5 MeV (e.g. Stecker \cite{stecker}; Dermer
\cite{dermer}) one expects the nucleonic $\gamma$-ray spectrum to have an
almost single power-law form to about 1 TeV, before turning over 
above 10~TeV. At the same time the leptonic Inverse Compton (IC)
$\gamma$-ray spectrum may be quenched at a lower cut-off energy due to
electron synchrotron losses (as in our model). Then $\gamma$-ray
observations within the range from 
10 MeV to 100~TeV would be useful in discriminating between the nucleonic
and leptonic components.

Recent observations of nonthermal X-rays and $\gamma$-rays indicate that
at least CR electrons are accelerated in SNRs. SN~1006 is one of the SNRs
for which there is evidence that electrons reach energies of about 100~TeV
(Koyama et al. \cite{koyama}; Tanimori et al. \cite{tan98}, \cite{tan01}).
It is also one of the three shell type SNRs in which TeV $\gamma$-ray
emission has been detected up to now. However, the interpretation of these
data is not unique. Depending on the assumed values for the unknown
physical parameters of SN~1006 (mainly the value of the magnetic field,
the electron to proton ratio, and to some extent also the nucleon
injection rate), the observed high-energy $\gamma$-ray emission can be
predominantly either inverse Compton radiation due to CR electrons
scattering on the microwave background radiation (as predicted 
by Mastichiadis (\cite{mas96}), Pohl (\cite{pohl}),
Mastichiadis \& de Jager (\cite{mast96}), and Yoshida \& Yanagita
(\cite{yoshida}) from the X-ray synchrotron emission, and through the
interpretation of the X-ray and $\gamma$-ray data by Aharonian
(\cite{ah99}), Aharonian \& Atoyan (\cite{aha99}), Berezhko et al.
(\cite{bkp99})), or $\pi^0$-decay emission due to hadronic
collisions of CRs with gas nuclei (Aharonian (\cite{ah99}), Aharonian \&
Atoyan (\cite{aha99}), Berezhko et al. (\cite{bkp99}), Berezhko
et al. (\cite{bkv01})).

The thermal and nonthermal X-ray emission has recently been
rediscussed by Allen et al. (\cite{allen}) who draw phenomenological
conclusions on the $\gamma$-ray emission to be dominated by IC radiation as
well as on the production and the characteristics of the nonthermal
nucleonic particle component which they estimate to have a total
energy of $10^{50}$~ergs.

In contrast, our starting point is the overall SNR dynamics and the
acceleration theory of the nonthermal component. Thus we accept the X-ray
results and the quantitative distinction between thermal X-ray emission
and the nonthermal synchrotron components, including the constraints on
the external density derived from them. We calculate the outer
shock size, speed and compression ratio as well as the nonthermal
quantities as solutions of the nonlinear, time-dependent kinetic equations
for electrons and protons as functions of space, time and particle
momentum, rather than phenomenologically assuming forms of the electron
and proton momentum distribution functions.

In this way we also obtain full morphological information. We determine
the unknown parameters like the upstream magnetic field strength and the
electron to proton ratio, and constrain the nucleon injection rate which
cannot yet be accuratetely calculated from known theory, by comparing with
the observations. For this purpose we use the selfconsistent kinetic model
of diffusive acceleration of CRs in SNRs (Berezhko et al. \cite{ber96};
Berezhko \& V\"olk \cite{bv97}) and investigate the $\gamma$-ray emission
from SN~1006. A preliminary version of our results is given in Berezhko et
al.  (\cite{bkv01}).

In contrast to a previous study (Berezhko et al. \cite{bkp99}) we restrict
ourselves to the so-called Bohm limit for CR diffusion near the shock,
assuming efficient and strong excitation of magnetohydrodynamic waves by
the accelerating particles themselves. This is consistent with the large
magnetic field strengths which we infer for this remnant.

Our considerations show that, together with a renormalization due to the
lack of spherical symmetry of the nucleon injection, the existing SNR data
are consistent with very efficient acceleration of CR nuclei at the SN
shock wave which converts a significant fraction of the initial SNR energy
content into CR energy as required for a typical source of the Galactic CR
nuclei. The relative amount of energetic electrons is rather small, and
may require additional sources to SNe type Ia like SN 1006. Therefore the
observed $\gamma$-ray emission of SN~1006 can indeed be of hadronic
origin, and we consider this as the physically most plausible solution
from the point of view of acceleration theory.

Nevertheless, the existing observations do not strongly exclude a solution
in which nuclear CRs play no important role and all nonthermal emissions
are of leptonic origin. Therefore we analyze whether the existing data can
also be fitted by an essentially different set of parameters. We
demonstrate that there is an alternative possibility to fit the overall
data, albeit with somewhat lower quality.

It implies a (physically not plausible) much lower injection rate and
acceleration efficiency of protons, a rather large electron to proton
ratio, and a lower magnetic field value compared with the case of
efficient CR nucleon acceleration. If we would assume SN~1006 to be a
typical representative, in this second case we could not consider the SNRs
as the source population of the Galactic CRs due to the low proton
acceleration efficiency. To discriminate empirically between these rather
different scenarios from our results $\gamma$-ray measurements above
10~TeV are required: in this very high energy range we expect a measurable
$\gamma$-ray flux only in the case of efficient nucleonic CR production.
We also confirm the earlier result (Aharonian \cite{ah99}; Aharonian \&
Atoyan \cite{aha99}; Berezhko et al. \cite{bkp99}; Berezhko et al.  
\cite{bkv01}) that the IC $\gamma$-ray emission should cover the SNR
almost uniformly, whereas the $\pi^0$-decay emission should be peaked
behind the shock front and limited to two polar caps where the
interstellar magnetic field is parallel to the shock normal.

\section{Model}

Since SN~1006 is a type Ia supernova (SN) we suggest that its evolution
takes place in a uniform interstellar medium (ISM), not modified by a wind
from the progenitor star. The general picture is then well-known.

A SN explosion ultimately ejects a shell of matter with total energy
$E_\mathrm{sn}$ and mass $M_\mathrm{ej}$. During an initial period the
shell material has a broad distribution in velocity $v$. The fastest part
of these ejecta is described by a power law $dM_\mathrm{ej}/dv\propto
v^\mathrm{2-k}$ (e.g. Jones et al. \cite{Jones81}; Chevalier \cite{chev81}).  
The interaction of the ejecta with the ISM creates a strong shock there
which accelerates particles.

Our nonlinear model (Berezhko et al. \cite{byk96};  Berezhko \& V\"olk
\cite{bv97}) is based on a fully time-dependent solution of the CR
transport equation together with the gas dynamic equations in spherical
symmetry. It yields at any given phase of the SNR evolution the complete
spatial distribution of gas and accelerated CRs and their spectrum. This
makes it possible to calculate any kind of emission, and its morphology,
produced by the accelerated particles.

The CR diffusion coefficient is taken as the Bohm limit 
\begin{equation}
\kappa (p)=\kappa (mc)(p/mc), \label{eq1}
\end{equation}
where approximately $\kappa(mc)=mc^2/(3eB)$, $e$ and $m$ are the particle
charge and mass, $p$ denotes the particle momentum, $B$ is the magnetic
field strength, and $c$ is the speed of light.  With this diffusion
coefficient we assume that the production of scattering waves is so strong
that limitations due to wave refraction (Malkov et al. \cite{maldj01}) do
not come in to influence the solution.

The number of suprathermal protons injected into the acceleration process
is described by a dimensionless injection parameter $\eta$ which is a
fixed fraction of the ISM particles entering the shock front. For
simplicity it is assumed that the injected particles have a velocity four
times higher than the postshock sound speed. Unfortunately there is no
complete selfconsistent theory of a collisionless shock transition which
can predict the value of the injection rate and its dependence on the
shock parameters. For the case of a purely parallel plane shock hybrid
simulations predict a quite high ion injection (e.g. Scholer et al.  
\cite{scholer}; Bennet \& Ellison \cite{benel}) which corresponds to the
value $\eta \sim 10^{-2}$ of our injection parameter. Such a high
injection is consistent with analytical models (Malkov \& V\"olk
\cite{malkv95}, \cite{malkv96}; Malkov \cite{malkov}) and confirmed by
measurements near the Earth's bow shock (Trattner \& Scholer \cite{trat}).
We note however that in our spherically symmetric model these results can
only be used with some important modification. In reality we must consider
the evolution of the large scale SN shock which expands into the ISM with
its magnetic field. In the case of SN~1006, at the current evolutionary
phase, the shock has a size of several parsecs. On such a scale the
unshocked interstellar magnetic field can be considered as uniform since
its random component is characterized by a much larger main scale of about
100~pc. Then our spherical shock is quasi-parallel in the polar regions
and quasi-perpendicular in the equatorial region. This magnetic field
essentially suppresses the leakage of suprathermal particles from the
downstream region back upstream when the shock is more and more oblique
(Ellison et al. \cite{elbj}; Malkov \& V\"olk \cite{malkv95}). Applied to
the spherical shock in the uniform external magnetic field it would mean
that only small regions near the poles, covering less than 10\% of the
shock surface, allow a sufficiently high injection that ultimately leads
to the transformation of an essential part (more than a few percent) of
the shock energy into CR energy, whereas the main part of the shock is an
inefficient CR accelerator. If one takes into account the Alfv\'{e}n wave
excitation due to CR streaming (which becomes efficient already at a very
low injection rate $\eta \sim 10^{-7}$) the local injection rate has to be
averaged over the fluctuating magnetic field directions and is lower than
for the purely parallel case by a factor of hundred. Therefore we adopt
here the value $\eta\sim 10^{-4}$ for the injection parameter (V\"olk et
al. \cite{voelk}). The detailed choice $\eta = 2\times 10^{-4}$ in the next
section is made in order to achieve a nonlinear shock modification that is
consistent with the observed radio spectral index.

According to our above estimate a substantial part of the shock still
efficiently injects and accelerates CRs. This fraction is effectively
increased relative to the above percentage due to broadening of the
injection region by the strong wave field as well as by CR diffusion
perpendicular to the mean magnetic field. In addition, the overall
conservation equations ensure an approximately spherical character of the
overall dynamics. Therefore, we assume the spherically symmetric approach
for the nonlinear particle acceleration process to be approximately valid
in those shock regions where injection is efficient. To take the effective
injection fraction $f_\mathrm{re}<1$ into account, we then need to
introduce a renormalization factor for the CR acceleration efficiency and
for all the effects which it produces in the SNR. A rough estimate
performed for the simple case of spherical shock expended in the uniform
outer magnetic field (V\"olk et al. \cite{voelk}) gives
$f_\mathrm{re}=0.15 ~{\rm to}~ 0.25$.

Note that such a picture is consistent with the observed structure of
SN~1006: the intense radio and X-ray emissions come from two bright rims
with radially oriented magnetic field (Reynolds \& Gilmore \cite{reng}).
They may be the polar regions of a quasispherical shock in an outer
magnetic field. The $\gamma$-ray observations may indicate a similar
asymmetry (Tanimori et al. \cite{tan98}) which would be well explained by
a corresponding concentration of the accelerated nucleons towards the
polar regions.

We assume that electrons are also injected into the acceleration process,
still at nonrelativistic energies below $m_\mathrm{e} c^2$. Since the
electron injection mechanism is not very well known (e.g. Malkov \& Drury
\cite{malkd}) for simplicity we consider their acceleration starting from
the same momentum as that of protons. At relativistic energies (for
protons) these electrons have exactly the same dynamics as the protons.
Therefore, neglecting synchrotron losses, 
their distribution function at any given time has the form
\begin{equation}
f_\mathrm{e}(p)= K_\mathrm{ep} f(p) \label{eq2}
\end{equation} 
with some factor $K_\mathrm{ep}$ which is about $10^{-2}$ for the
average CRs in 
the Galaxy. We take it here as a parameter.

The electron distribution function
$f_\mathrm{e}(p)$ deviates only at sufficiently large momenta from
this relation 
due to synchrotron losses,
which are taken into account by supplementing the ordinary diffusive
transport equation with a loss term:
\begin{equation}
{\partial f_\mathrm{e}\over \partial t}=\nabla \kappa \nabla
f_\mathrm{e} 
-\vec{w}\nabla f_\mathrm{e}
+\frac{\nabla \vec{w}}{3}p\frac{\partial f_\mathrm{e}}{\partial p}
-\frac{1}{p^2}\frac{\partial}{\partial p}
\left( \frac{p^3}{\tau_\mathrm{l} }f_\mathrm{e}\right), \label{eq3}
\end{equation}
where the first three terms in the right hand side of this equation
describe diffusion, convection due to the mass velocity $\vec{w}$ of 
the gas 
and adiabatic effects, respectively. The synchrotron loss time in the
third term is determined by the expression (e.g. Berezinskii et al. 
\cite{brz90})
\begin{equation}
\tau_\mathrm{l}=\left( \frac{4 r_0^2 B^2p}{9
    m_\mathrm{e}^2c^2}\right)^{-1}, 
\label{eq4}
\end{equation}
where $m_\mathrm{e}$ is the electron mass and
$r_0$ the classical electron radius.

The solution of the dynamic equations at each instant of time yields the
CR spectrum and the spatial distributions of CRs and gas. This allows us
to calculate the expected flux $F_{\gamma}^\mathrm{\pi}(\epsilon_{
\gamma})$ of $\gamma$-rays from $\pi^0$-decay due to hadronic (p-p)  
collisions of CRs with the gas nuclei. Following the work of Dermer 
(\cite{dermer})
and its later improvement by Naito \& Takahara 
(\cite{naitot}) we use here the
isobar model at the protons kinetic energies $\epsilon_k<3$~GeV 
and the scaling model at
$\epsilon_k>7$~GeV with a linear connection between 3 and 7~GeV. 
This model agrees very well with the simpler
approach, introduced by Drury et al. (\cite{dav94}) (see also Berezhko \&
V\"olk \cite{bv97}, \cite{bv00}; Berezhko et al. \cite{bkp99}) at high
energies $\epsilon_{\gamma}>0.1$~GeV, except in the cutoff region, where
the scaling model yields an essentially smoother turnover of the
$\gamma$-ray spectrum at somewhat lower energies.

The choice of $K_\mathrm{ep}$ allows us to determine in addition the electron 
distribution function and
to calculate the associated emission. The expected synchrotron 
flux at distance $d$
from the SNR is given by the expression (e.g. Berezinskii et al.
\cite{brz90})
\begin{equation}
S_{\nu}={3\times 10^{- 21}\over d^2}
\int_0^{\infty} dr r^2 B_{\perp}\int_0^{\infty} dp p^2 f_\mathrm{e}(r,p) 
F\left(\frac{\nu}{\nu_\mathrm{c}} \right)  \label{eq5}
\end{equation}
in erg/(cm$^2$s), where
\[
F(x)=x\int_x^{\infty}K_{5/3}(x')dx',
\]
$K_{\mu}(x)$ is the modified
Bessel function, 
$\nu_\mathrm{c}=3e B_{\perp} p^2 /[ 4\pi (m_\mathrm{e}c)^3]$ and
$B_{\perp}$ is the  magnetic field component
perpendicular to the line of sight.

Since in the shock region where particle injection and acceleration is
efficient the upstream magnetic field is assumed to be almost completely
randomized due to intense Alfv\'{e}n wave generation. For such a strongly
turbulent upstream field, consistent with Bohm diffusion, the postshock
field strength $B_2$ is given by by $B_2=\sigma B_0$, where 
$\sigma^2_B = 1/3 + (2/3)\sigma^2$. We shall approximately equate
$\sigma_B$ with the gas compression ratio $\sigma$ at the shock that
We suggest also that in such a young SNR like SN~1006 the downstream
magnetic field $B_\mathrm{d}$ in a relatively thin region between the
shock and the ejecta is approximately uniform and we shall take
$B_\mathrm{d}=B(r<R_\mathrm{s})=B_2$.

The relativistic electrons produce $\gamma$-ray emission due to
IC scattering of background photons. It is not
difficult to show that due to the hard spectrum of
accelerated electrons only the 2.7 K cosmic microwave background (CMB) is
important in the case considered. The expected differential flux of IC
$\gamma$-rays as a function of their energy $\epsilon_{\gamma}$ can be
represented in the form:
\[
{dF_{\gamma}^\mathrm{IC}\over d\epsilon_{\gamma}}=
{4\pi  c \over d^2}
\int_0^{\infty} dr r^2 
\int_0^{\infty} d\epsilon n_\mathrm{ph}(\epsilon)
\]
\begin{equation}
\hspace{1.5cm} \times \int_{p_\mathrm{min}}^{\infty} dp p^2
\sigma(\epsilon_\mathrm{e},
\epsilon_{\gamma},\epsilon)f_\mathrm{e}(r,p) 
\label{eq6}
\end{equation}
in photons/(cm$^2$s erg),
where (Blumenthal \& Gould \cite{blgo})
\[
\sigma(\epsilon_\mathrm{e},\epsilon_{\gamma},\epsilon)=
\frac{3\sigma_\mathrm{T}(m_\mathrm{e}c^2)^2}{4\epsilon
\epsilon_\mathrm{e}^2}
\]
\begin{equation}
\hspace{1cm} \times \left[ 
2q\ln q+(1+2q)(1-q)+0.5\frac{(\Gamma q)^2(1-q)}{1+\Gamma  q}
\right]         \label{eq7}
\end{equation}
is the differential cross section for the up-scattering of a photon
with incident energy $\epsilon$ to energy $\epsilon_{\gamma}$ by
the elastic collision with an electron of energy $\epsilon_\mathrm{e}$,
\begin{equation}
n_\mathrm{ph}=\frac{1}{\pi^2 (\-\hbar c)^3}
\frac{\epsilon ^2}{\exp(\epsilon/k_\mathrm{B}T)-1}         
\label{eq8}
\end{equation}
is the blackbody spectrum of the CMB, $\hbar$ and
$k_B$ are Planck and Boltzmann constant respectively, $T=2.7$~K,
$\sigma_\mathrm{T}=6.65 \times 10^{-25}$ cm$^2$ is the Thomson
cross-section, $q=\epsilon_{\gamma}/[\Gamma (\epsilon_\mathrm{e}
-\epsilon_{\gamma})]$,
$\Gamma = 4\epsilon \epsilon_\mathrm{e} /(m_\mathrm{e}c^2)^2$ 
, and $p_\mathrm{min}$ is the minimal momentum of electron, whose
energy $\epsilon_\mathrm{e}$ is determined by the condition $q=1$.

Energetic electrons produce also nonthermal bremstrahlung (NB) emission,
interacting with both ambient electrons and protons. We use standard
formulae for the cross section to calculate the NB $\gamma$-ray flux
$F^\mathrm{NB}_{\gamma}(\epsilon_{\gamma})$ 
(e.g. Baring et al. \cite{bare}). As
it was already demonstrated by these last authors, in young SNRs the
primary energetic electron number density due to electrons directly
accelerated from the suprathermal postshock distribution by far exceeds
that of secondaries which were produced via the decay of charged pions
created in p--p collisions. Hence the contribution of secondaries to the
NB, IC, and synchrotron emission can be neglected.

\section{Results and Discussion}

Since SN~1006 is a type Ia SN we use typical SN Ia parameters 
in our calculations:
ejected mass $M_\mathrm{ej}=1.4 M_{\odot}$, $k=7$, and
a uniform ambient ISM. 

The ISM density is the most relevant parameter, especially for the
$\gamma$-ray production (e.g. Berezhko et al. \cite{bkp99}). Here we use
the most appropriate value of the ambient number density of ISM hydrogen
$N_\mathrm{H}=0.3$~cm$^{-3}$ as well as the distance $d=1.8$~kpc,
consistent with X-ray and optical imagery of SN~1006 (Winkler \& Long
\cite{wl97}, see also Allen et al. \cite{allen} and references therein).

The value $T_0=10^4$~K is used for the ISM temperature. Note that SNR
and CR dynamics are not sensitive to $T_0$, because the shock
structure is mainly determined by the Alfv\'{e}nic Mach number.

The other important parameter is the ISM magnetic field $B_0$.
Unfortunately there is no direct way to determine its value from the
observations. Since it influences the synchrotron spectrum in an essential
way, we use an upstream magnetic field value $B_0=20$~$\mu$G. It is
required to provide the observed shape of the synchrotron spectrum in the
radio and the X-ray bands.

The gas dynamics problem is characterized by the following length, 
time, and velocity scales: 
\[R_0=(3 M_\mathrm{ej}/ 4\pi \rho_0)^{1/3},~ t_0=R_0/V_0,~
V_0=\sqrt{2E_\mathrm{sn}/{M_\mathrm{ej}}},
\]
which are the sweep-up radius, sweep-up time and mean ejecta speed
respectively. Here  $\rho_0=1.4 m_\mathrm{p}N_\mathrm{H}$ is the ISM
mass density; $m_\mathrm{p}$ is the proton mass. 

The
shock expansion law during the free expansion phase ($t<t_0$) is then
\begin{equation}
R_\mathrm{s} \propto E_\mathrm{sn}^\mathrm{(k-3)/2k} \rho_0^\mathrm{-1/k}
t^\mathrm{(k-3)/(k-2)} 
\label{eq9}
\end{equation}
(Chevalier \cite{chev81}) which for $k=7$ gives
\begin{equation}
R_\mathrm{s}\propto (E_\mathrm{sn}^2/ \rho_0)^{1/7}t^{4/5}. \label{eq10}
\end{equation}
In the adiabatic phase $(t\,\,\raisebox{0.2em}{$>$}\!\!\!\!\!
\raisebox{-0.25em}{$\sim$}\,\, t_0)$ we have 
\begin{equation}
R_\mathrm{s}\propto (E_\mathrm{sn}/
\rho_0)^{1/5}t^{2/5}. \label{eq11}
\end{equation}

The observed expansion law of SN~1006 (Moffett et al.\ \cite{mof93}) is 
$R_\mathrm{s}\propto t^{\mu}$
with $\mu= 0.48 \pm 0.13$. We conclude that within the 
observational errors SN~1006 
should already be in the adiabatic phase.

\begin{figure}
\centering
\includegraphics[width=7.5cm]{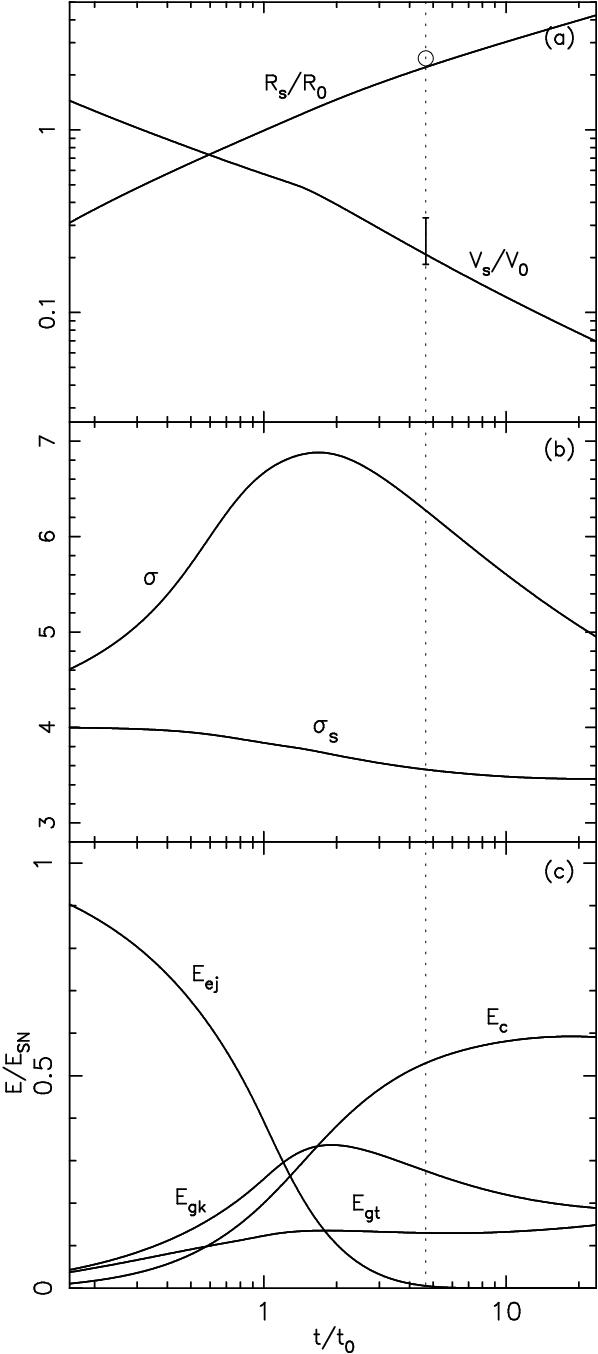}
\caption{(a) Shock radius $R_\mathrm{s}$ and shock speed
  $V_\mathrm{s}$; 
(b) total shock 
($\sigma$)
and subshock ($\sigma_\mathrm{s}$) compression ratios; (c) ejecta 
($E_\mathrm{ej}$), 
CR
($E_\mathrm{c}$), gas thermal ($E_\mathrm{gt}$), and gas kinetic 
($E_\mathrm{gk}$) energies
as a function of time.
Scale values are $R_0=3.2$~pc, $V_0=14 675$~km/s,
$t_0=212$~years. The dotted vertical line marks the current 
epoch. 
The observed size and speed of the shock
(Moffett et al.\ \cite{mof93}), are shown as well.} 
\label{f1}
\end{figure}

\begin{figure} 
\centering 
\includegraphics[width=7.5cm]{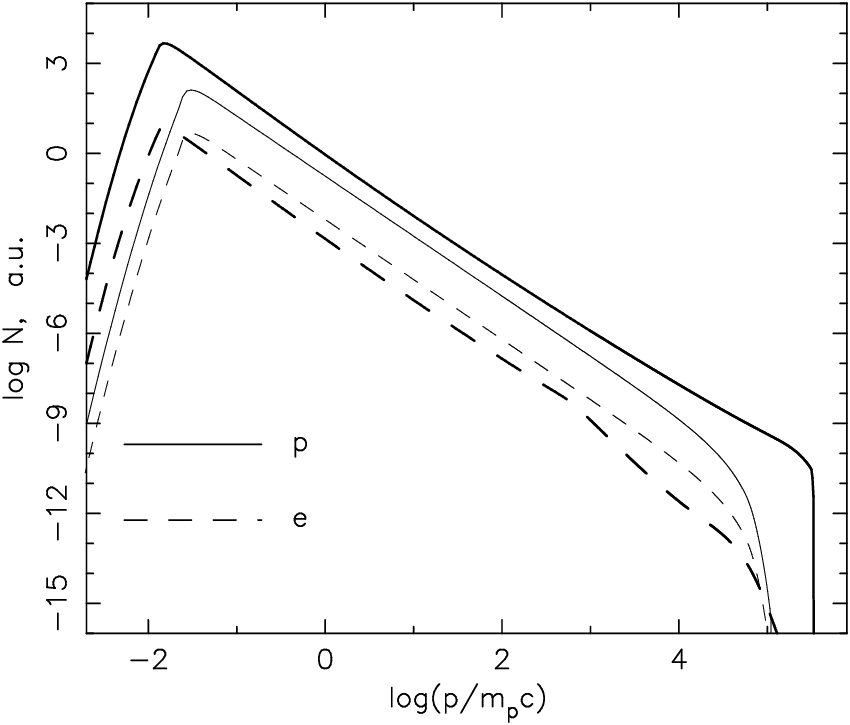}
\caption{The overall (= spatially integrated) CR spectrum as function of
momentum. Solid and dashed
lines correspond to protons and electrons, respectively. Thick and thin
lines correspond to efficient and inefficient proton acceleration,
respectively.} 
\label{f2} 
\end{figure}

\begin{figure*}
\sidecaption
\includegraphics[width=12cm]{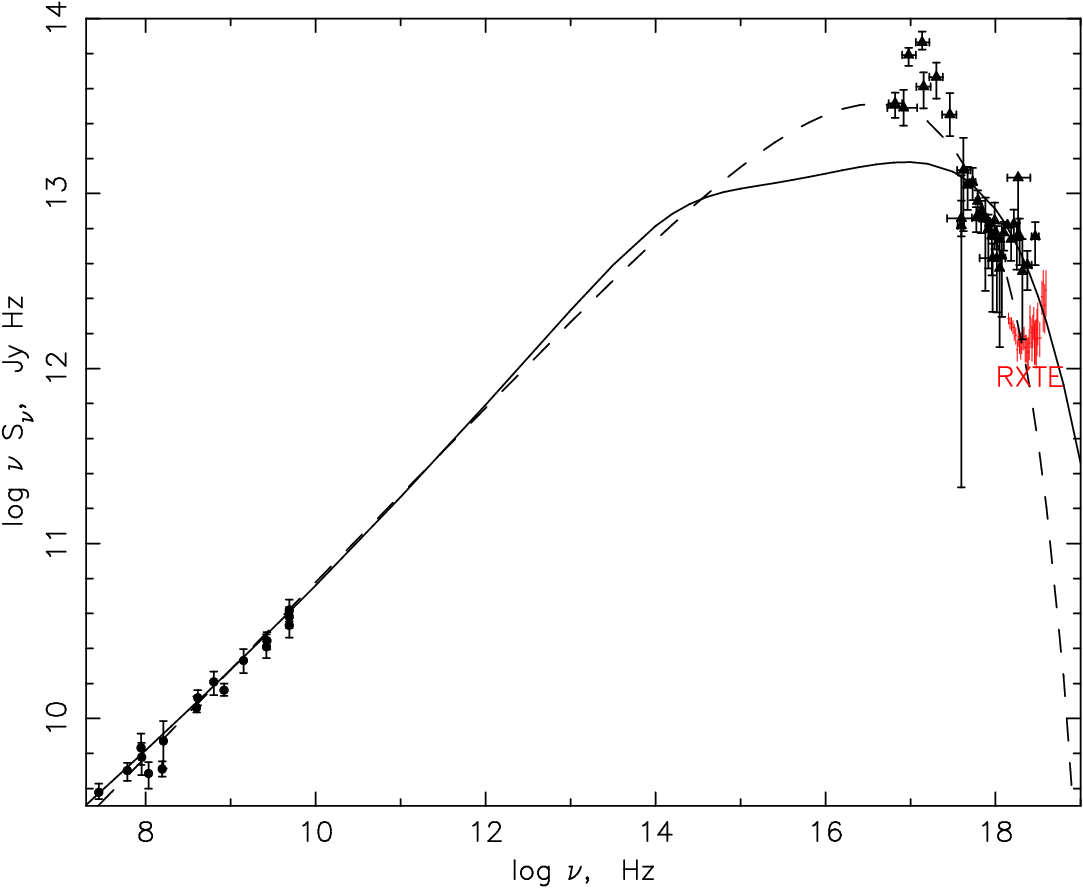}
\caption{Synchrotron emission flux as a function of frequency for the
same two cases as in Fig.\ref{f2}. Solid and dashed
lines correspond to efficient and inefficient proton acceleration,
respectively. The observed
X-ray (in black color: Hamilton et al. \cite{ham86}; in red color: Allen
et al. \cite{allen1}) 
and radio emissions (in black color: Reynolds \cite{reyn96}) are shown.}
\label{f3}
\end{figure*}

\begin{figure}
\centering
\includegraphics[width=7.5cm]{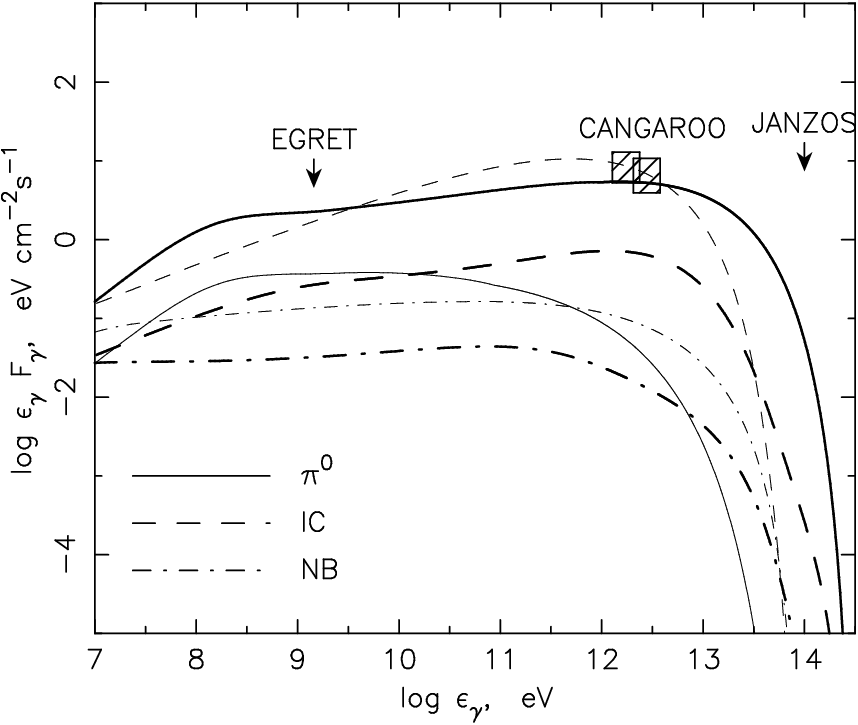}
\caption{IC (dashed lines), $\pi^0$-decay (solid lines) and NB (dash-dotted
lines) integral
$\gamma$-ray energy 
fluxes as a function of $\gamma$-ray energy for the same cases as in 
Fig.\ref{f2}. 
High energy $\gamma$-ray data (Tanimori et al. \cite{tan98}), the
EGRET upper limit as given by Mastichiadis \& de Jager \cite{mast96}, and
the JANZOS upper
limit (Allen et al. \cite{allenwh}) are shown.}
\label{f4}
\end{figure}

\begin{figure}
\centering
\includegraphics[width=7.5cm]{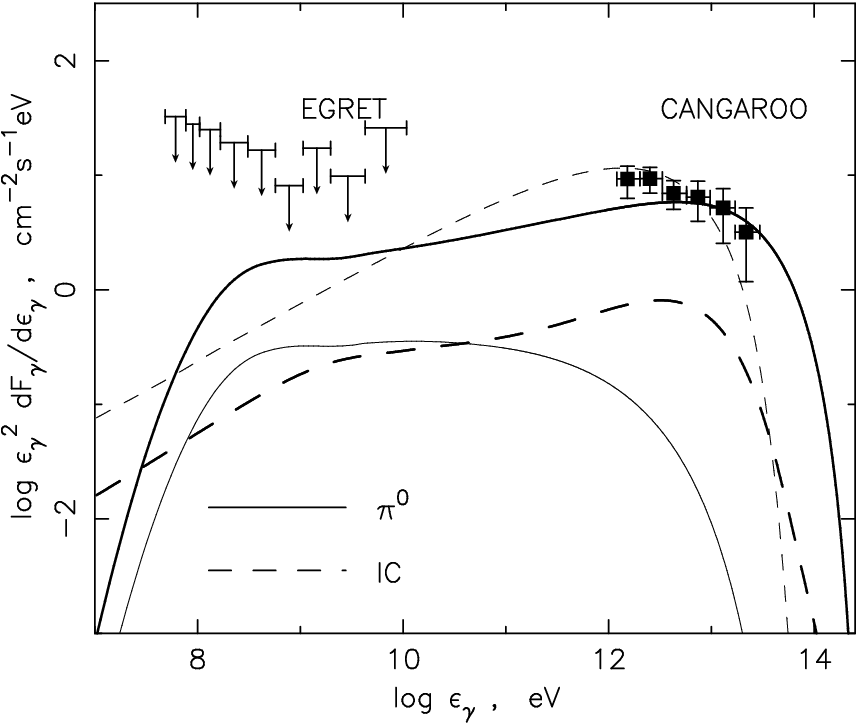}
\caption{Differential $\pi^0$-decay ({\it solid lines}) and IC ({\it 
dashed lines})
$\gamma$-ray energy fluxes as a function of $\gamma$-ray energy 
for the same cases as in 
Fig.\ref{f2}. The recent differential
high energy $\gamma$-ray energy flux  data (Tanimori et al. \cite{tan01}) 
and
EGRET upper limits (Naito et al. \cite{naitoetal}) are also shown.}
\label{f5}
\end{figure}

The calculations together with the experimental data are shown in
Fig.\ref{f1}--\ref{f5}. For the assumed SN distance and ISM density an
explosion energy $E_\mathrm{sn}=3\times 10^{51}$~erg fits the observed SNR
size $R_\mathrm{s}$ and its expansion rate $V_\mathrm{s}$ (Moffet et al.
\cite{mof93}). The shown uncertainties of $R_\mathrm{s}$ and 
$V_\mathrm{s}$ are calculated
as the uncertainties of the angular size and expansion rate times the
distance $d=1.8$~kpc.

According to Fig.\ref{f1}a SN~1006 is indeed already in the
adiabatic phase. The assumed injection rate $\eta=2\times 10^{-4}$
leads to a significant modification of the shock which at the current
epoch, $t=995$~yr, has a total compression ratio $\sigma=6.3$ and a
subshock compression ratio $\sigma_\mathrm{s}=3.6$ (Fig.\ref{f1}b).

The different SNR energy components, the ejecta energy $E_\mathrm{ej}$,
the gas kinetic ($E_\mathrm{gk}$), and thermal ($E_\mathrm{gt}$) energies,
and the CR energy $E_\mathrm{c}$, are presented in Fig.\ref{f1}c as
functions of time. The acceleration process is characterized by a high
efficiency in spherical symmetry:  at the current time $t/t_0=4.68$ about
53\% of the explosion energy has been already transferred to CRs, and the
CR energy content $E_\mathrm{c}$ continues to increase to a maximum of
about 60\% in the later Sedov phase (Fig.\ref{f1}c), when particles start
to leave the source. As usually predicted by the model, the CR
acceleration efficiency is significantly higher than required for the
average replenishment of the Galactic CRs by SNRs, corresponding to
$E_\mathrm{c}\approx 0.1E_\mathrm{sn}$.  As discussed before, this
discrepancy can be attributed to the physical conditions at the shock
surface which influence the injection efficiency. The magnetic field
geometry is the most important factor: at the quasiperpendicular portion
of the shock ion injection (and subsequent acceleration) is presumably
depressed compared with the quasiparallel portion. Therefore the number of
nuclear CRs, calculated within the spherically-symmetrical approximation,
should be renormalized by this depression factor (see V\"olk et al.
\cite{voelk} for details). Assuming SN 1006 to be an average Galactic CR
source, the renormalizing factor should be $f_\mathrm{re}=0.2$. It means
that according to our calculation the actual energy of accelerated protons
in SN~1006 at the present epoch is
\begin{equation}
E_\mathrm{c}=0.53f_\mathrm{re}E_\mathrm{sn}= 3.2\times
10^{50}~\mbox{erg}. 
\label{eq12}
\end{equation}
Note that such a factor is consistent with observations which show that
efficient CR acceleration takes place at about 25\% of the shock surface
(e.g. Allen et al. \cite{allen}).

The volume-integrated (or overall) CR spectrum 
 \begin{equation}
N(p,t)=16\pi^2p^2 \int_0^{\infty}dr r^2 f(r,p,t)  
\label{eq13}
\end{equation} 
has, for the case of protons, almost a pure power-law form $N\propto
p^{-\gamma}$ over a wide momentum range from $10^{-2}m_\mathrm{p}c$ up to
the cutoff momentum $p_\mathrm{max}=\epsilon_\mathrm{max}/c$, where
$\epsilon_\mathrm{max}\approx 3\times 10^{14}$~eV is the maximum CR energy
(Fig.\ref{f2}). The limitation of this numerically determined value
$p_\mathrm{max}$ can be understood mainly in terms of the influence of
geometrical factors which are the finite size and speed of the shock, its
deceleration and the adiabatic cooling effect in the downstream region
(Berezhko \cite{ber96}). Due to the shock modification the power-law index
slowly varies from $\gamma=2.1$ at $p\,\,\raisebox{0.2em}{$<$}\!\!\!\!\!
\raisebox{-0.25em}{$\sim$}\,\, m_\mathrm{p}c$ to $\gamma=1.8$ at
$p\,\,\raisebox{0.2em}{$>$}\!\!\!\!\!
\raisebox{-0.25em}{$\sim$}\,\,10^3m_\mathrm{p}c$.

The shape of the overall electron spectrum $N_\mathrm{e}(p)$ deviates from
that of the proton spectrum $N(p)$ at high momenta $p>p_\mathrm{l}\approx
10^3m_\mathrm{p}c$, due to the synchrotron losses in the downstream region
with magnetic field $B_\mathrm{d}\approx 120~\mu$G which is assumed
uniform in this region ($B_\mathrm{d} \approx B_2=\sigma B_0$). According
to expression (\ref{eq4}) the synchrotron losses should become important
for electron momenta greater than
\begin{equation} 
\frac{p_\mathrm{l}}{m_\mathrm{p}c} \approx 
1.3 \left(\frac{10^8~\mbox{yr}}{t}\right)
\left(\frac{10~\mu\mbox{G}}{B_\mathrm{d}}\right)^2. \label{eq14}
\end{equation}
Substituting the SN age $t=10^3$~yr into this expression, we have
$p_\mathrm{l}\approx 600m_\mathrm{p}c$, in reasonable agreement with the 
numerical results
(Fig.\ref{f2}).

In detail the momentum dependence of the spatially integrated electron
spectrum $N_e$ from Eq. (13) is somewhat more complicated at high
energies. This can be understood as follows: 

The shock constantly produces the electron spectrum $f_\mathrm{e}\propto
p^{-q}$ with $q \approx 4$ up to the maximum momentum
$p_\mathrm{max}^\mathrm{e}$ which is much larger than $p_\mathrm{l}$. This
spectrum is not modified by synchrotron losses within the downstream
region of thickness
\begin{equation}
l=\tau_\mathrm{l} u_2, \label{l}
\end{equation}
where $u_2=V_\mathrm{s}/\sigma$ is the gas speed with respect to the
shock. At low momenta $p<p_\mathrm{l}$ the loss time $\tau_\mathrm{l}$
exceeds the age of the system. Formally this means that the size of the
lossfree region is greater than the thickness $\Delta r\approx
R_\mathrm{s}/(3\sigma)$ of the swept up shell. At large momenta $p>p_\mathrm{l}$ 
we have $l<\Delta r$. Within the downstream region
$r<R_\mathrm{s}-l$ the electron spectrum therefore becomes essentially
modified ($f_\mathrm{e}\propto p^{-5}$) by the influence of the
synchrotron losses. At a given radius $r<R_\mathrm{s}$ accelerated
electrons have an age of $\Delta t\approx (R_\mathrm{s}-r)/u_2$ and
therefore, cf. Eq. (14), their spectrum has a break at
$p_\mathrm{l}(r)=p_\mathrm{l} \times (t/\Delta t)$. According to its
definition (\ref{eq13}) the spatially integrated electron spectrum
$N_\mathrm{e}(p)$ is a mixture of electron spectra with different
$p_\mathrm{l}(r)$. Qualitatively one would expect that within the momentum
range $p_\mathrm{l}~\mathrm{to}~p_\mathrm{max}^\mathrm{e}$, the overall
electron spectrum is close to the form $N_\mathrm{e}\propto p^{-3}$, 
somewhat
hardening towards $p_\mathrm{max}^\mathrm{e}$. This is in agreement with
the calculation. 

The maximum electron momentum $p_\mathrm{max}^\mathrm{e}$ can be estimated
by equating the synchrotron loss time (\ref{eq4}) and the acceleration
time
\begin{equation}
\tau_\mathrm{a}=\frac{3}{\Delta u}\left(
\frac{\kappa_1}{u_1}+\frac{\kappa_2}{u_2}
\right), \label{eq15}
\end{equation}
where $u_1=V_\mathrm{s}$ and $u_2=u_1/\sigma$ are the upstream and
downstream gas velocities relative to the shock front, and $\Delta
u=u_1-u_2$. During the acceleration time $\tau_\mathrm{a}$ the electrons
spent the time $3\kappa_1/(u_1\Delta u)$ in the upstream magnetic field
$B_0$, and $3\kappa_2/(u_2\Delta u)$ in the downstream magnetic field
$B_2=\sigma_\mathrm{B} B_\mathrm{0}$ (in the case considered
$\sigma_\mathrm{B}=\sigma$). Therefore, in the general case the maximum
electron momentum $p^\mathrm{e}_\mathrm{max}$ is a solution of the
equation
\begin{equation}
\tau_\mathrm{a}(p)^{-1}=[\tau_\mathrm{l}^{-1}(B_0,p)+
\tau_\mathrm{l}^{-1}(B_2,p)\sigma/\sigma_\mathrm{B}]
/(1+\sigma/\sigma_\mathrm{B}), \label{eq16}
\end{equation}
which can be written in the form
\[
\frac{p_\mathrm{max}^\mathrm{e}}{m_\mathrm{p}c}
= 6.7\times 10^4 \left(\frac{V_\mathrm{s}}{10^3~\mbox{km/s}}\right)
\]
\begin{equation}
\hspace{1cm}\times
\sqrt{\frac{(\sigma-1)}{\sigma (1+\sigma_\mathrm{B} \sigma)}
  \left(\frac{10~\mu\mbox{G}}{B_0}\right)}. \label{eq17}
\end{equation}
The main part of the electrons with the highest energies
$\epsilon_\mathrm{e}\,\,\raisebox{0.2em}{$>$}\!\!\!\!\!
\raisebox{-0.25em}{$\sim$}\,\, 10$~TeV is produced at the end of the free
expansion phase. At this stage $V_\mathrm{s}\sim V_0$ which leads to a
maximum electron momentum $p_\mathrm{max}^\mathrm{e}\approx 4\times
10^4m_\mathrm{p}c$ in agreement with the numerical results (Fig.\ref{f2}).

The parameters $K_\mathrm{ep}=3\times 10^{-4}$ and $B_{\perp} \approx
0.3B_\mathrm{d}\approx 36$~$\mu$G provide good agreement between the
calculated and the measured synchrotron emission in the radio- and X-ray
ranges (Fig.\ref{f3}). The steepening of the electron spectrum at high
energies due to synchrotron losses naturally yields a fit to the X-ray
data with their soft spectrum. Note that two kinds of X-ray data are
presented in Fig.\ref{f3}. The Hamilton et al. (\cite{ham86})  data
represent the total (i.e. thermal plus nonthermal) X-ray emission in a
relatively wide frequency range, where the nonthermal synchrotron emission
contribution dominates at frequencies $\nu
\,\,\raisebox{0.2em}{$>$}\!\!\!\!\! \raisebox{-0.25em}{$\sim$}\,\,
10^{18}$~Hz. At these highest frequencies the Allen et al.  
(\cite{allen1}) RXTE nonthermal data are also presented (in red color), lying
slightly below the Hamilton et al. data points. Clearly our nonthermal
model spectra must lie below the Hamilton et al. spectral points at $\nu
\,\,\raisebox{0.2em}{$<$}\!\!\!\!\! \raisebox{-0.25em}{$\sim$}\,\,
10^{18}$~Hz. Such a smooth spectral behavior is achieved in a $20$~$\mu$G
upstream field.

Since the total number of accelerated protons has to be reduced by the
factor $f_\mathrm{re}=0.2$, the same number of accelerated electrons then
corresponds to a renormalized parameter $K_\mathrm{ep}=1.5\times 10^{-3}$.
Then the total energy of accelerated electrons in SN~1006 is
$E_\mathrm{c}^\mathrm{e}=5\times 10^{47}$~erg.

The ratio of accelerated electrons to protons thus comes out lower by a
factor of about 6 than the canonical ratio of 0.01, generally deduced from
observations of the Galactic CRs. This is an interesting consequence of
our model: if SN 1006 is typical for the nuclear Galactic CR source
population, then other sources like young Pulsars or Pulsar Nebulae must
also significantly contribute to the Galactic CR electron population (e.g.
Aharonian et al. \cite{aav}; Pohl \& Esposito \cite{pesp}). An additional
fraction might come from secondary interstellar electrons/positrons due to
the decay of charged pions that are produced in inelastic collisions of CR
nuclei with nuclei of the interstellar gas. Since the majority of Galactic
Supernovae is not of the type Ia to which SN 1006 belongs, but rather
occurs as a consequence of core collapse events, there is the alternative
possibility that core collapse Supernovae produce a higher electron to
proton ratio. Such an increase might have to do with the different
circumstellar magnetic field structure of massive stars, but this is
indeed only a possibility for which we have no proof at present.

The radio data are fitted by a power law spectrum $S_{\nu}\propto
\nu^{-\alpha}$, whose index $\alpha =0.57\pm 0.06$ (Allen et al.
\cite{allen}) is noticeably larger than 0.5, corresponding to the electron
spectrum $N_\mathrm{e}\propto p^{-2}$ produced by an unmodified shock with
compression ratio $\sigma=4$. In our case the shock is essentially
modified by the backreaction of the accelerated protons (see
Fig.\ref{f1}b): its compression ratio $\sigma=6.3$. At the same time low
energy electrons with momenta $p\,\,\raisebox{0.2em}{$<$}\!\!\!\!\!
\raisebox{-0.25em}{$\sim$}\,\, 10m_\mathrm{p}c$
($\epsilon_\mathrm{e}\,\,\raisebox{0.2em}{$<$}\!\!\!\!\!
\raisebox{-0.25em}{$\sim$}\,\, 10$~GeV), which produce synchrotron
emission at $\nu\,\,\raisebox{0.2em}{$<$}\!\!\!\!\!
\raisebox{-0.25em}{$\sim$}\,\, 10$~GHz, are accelerated at the {\it
subshock} which has the compression ratio $\sigma_\mathrm{s}=3.6$.  
Therefore these electrons have a steeper spectrum $N_\mathrm{e}\propto
p^{-2.1}$ that leads to the expected radio spectrum $S_{\nu}\propto
\nu^{-0.54}$, fitting the experimental data very well (Fig.\ref{f3}).  
The fact that the observed value of the radio power law index $\alpha$
exceeds the value 0.5 can be considered as an indication that the shock is
essentially modified. A relatively high upstream magnetic field strength
$B_0=20$~$\mu$G, compared with typical ISM values $B_0=5$~$\mu$G, and a
corresponding downstream value $B_{\perp}=36$~$\mu$G are required to have
radio emitting electron energies in the steep part of their spectrum
$\epsilon_\mathrm{e}\,\,\raisebox{0.2em}{$<$}\!\!\!\!\!
\raisebox{-0.25em}{$\sim$}\,\, 1$~GeV on the one hand, and to give a
smooth rollover of the synchrotron spectrum $S_{\nu}(\nu)$ at frequencies
$\nu=10^{15}~ \mathrm{to}~ 10^{18}$~Hz on the other.  According to model
calculations by Lucek and Bell (\cite{lucb00}), the existing ISM field can
indeed be significantly amplified near a strong shock by CR streaming.

Such a strongly nonlinear behavior has three consequences in that (i) it
leads to the high magnetic field strength, required to understand the
measured nonthermal synchrotron spectrum, (ii) it is consistent with our
assumption of Bohm diffusion, and (iii) it suggests a value $\eta \sim
10^{-4}$ for the injection parameter that we use (see also section 2).

The calculated integral IC and $\pi^0$-decay $\gamma$-ray energy fluxes 
are presented in
Fig.\ref{f4} and \ref{f5} together with the available experimental data.

In Fig.4 we present also the integral NB $\gamma$-ray fluxes; they play no
important role for SN~1006. Note that the number of accelerated protons,
which produce $\pi^0$-decay $\gamma$-rays, has been reduced by the factor
$f_\mathrm{re}=0.2$.

According to the calculation, the hadronic $\gamma$-ray production exceeds
the electron contribution by a factor of about 7 at energies
$\epsilon_{\gamma}\,\,\raisebox{0.2em}{$<$}\!\!\!\!\!
\raisebox{-0.25em}{$\sim$}\,\, 1$~TeV, and dominates at
$\epsilon_{\gamma}>10$~TeV (Fig.\ref{f4}). The calculation is in good
agreement with the TeV-measurements reported by the CANGAROO collaboration
(Tanimori et al.\ \cite{tan98}), and it does not contradict the EGRET
upper limit $F_{\gamma}^\mathrm{E}=8\times 10^{-9}$ cm$^{- 2}$s$^{-1}$ at
$\epsilon_{\gamma}=1.4$~GeV (cf. Mastichiadis \& de Jager \cite{mast96}).
This is also confirmed by Fig.\ref{f5}, where we compare our calculations
with the revised CANGAROO data (Tanimori et al. \cite{tan01}) plus an
extended set of EGRET upper limits (Naito et al. \cite{naitoetal}).  

The differential $\pi^0$-decay $\gamma$-ray flux
$dF_{\gamma}^{\pi}/d\epsilon_{\gamma}$ has the expected peak at photon
energy $m_{\pi}/2=67.5$~MeV (see also Naito \& Takahara \cite{naitot}).
This feature may be used for the identification of a hadronic contribution
to the observed $\gamma$-ray flux at the particle energies $< 10$~GeV 
mainly producing it.  However, two considerations weaken this possibility. 
First
of all, for reasons of instrumental sensitivity, we expect only nearby
SNRs to be detectable in the foreseeable future. Such objects have
diameters approaching 1 degree, and SN 1006 is an example. The diffuse
Galactic $\gamma$-ray background is large at these energies due to the
steep spectrum of the Galactic CRs that produce it, as pointed out by
Drury et al. (\cite{dav94}). Therefore, we expect SNRs to be best
observable at much higher $\gamma$-ray energies $\epsilon_{\gamma}
\,\,\raisebox{0.2em}{$>$}\!\!\!\!\!
\raisebox{-0.25em}{$\sim$}\,\, 100$~GeV, where the
hard source spectra typically dominate over the background emission.
Secondly, from a strictly empirical point of view, such a nuclear CR
component that produces the $\pi^0$-decay bump need not extend far beyond
the GeV region. For the existence, in a SNR, of a nuclear CR component
approaching the energy region of the knee in the Galactic spectrum, the
detection of a 67.5 MeV $\gamma$-ray feature is therefore neither a
necessary nor a sufficient condition.

On the other hand, due to the high magnetic field strength which we deduce
for this remnant, the $\gamma$-ray spectra produced by the electronic and
hadronic CR components have rather similar shapes in the energy interval
0.1~GeV$\,\,\raisebox{0.2em}{$<$}\!\!\!\!\! \raisebox{-0.25em}{$\sim$}\,\,
\epsilon_{\gamma}\,\,\raisebox{0.2em}{$<$}\!\!\!\!\!
\raisebox{-0.25em}{$\sim$}\,\,$10~TeV due to electron synchrotron losses.
As a consequence the best observational possibility to discriminate
between leptonic and hadronic contributions {\it at high energies} $\gg
1~$ GeV is to measure the $\gamma$-ray spectrum at energies higher than
10~TeV, where we expect these two spectra to be essentially different.
Therefore the clear detection of a substantial flux at energies
$\epsilon_{\gamma}\,\,\raisebox{0.2em}{$>$}\!\!\!\!\!
\raisebox{-0.25em}{$\sim$}\,\, 10$~TeV would provide direct evidence for
its hadronic origin.

\section{Inefficient proton acceleration model}
The set of parameters which was discussed until now is limited by the
existing measurements. There is almost no freedom for a substantial change
of any parameter value without a loss of agreement between theory and
experiment. For example, the proton injection rate which determines the
number of accelerated protons and the $\pi^0$-decay $\gamma$-ray emission,
was taken to produce the shock modification which in turn is required to
produce a steep electron spectrum $N_\mathrm{e}\propto p^{-2.1}$ needed to
fit the observed radio emission of SN~1006. For this injection rate
$\eta=2\times 10^{-4}$ 
%
%
a relatively high upstream magnetic field is unavoidably required in order
to produce the observed synchrotron emission.

On the other hand one may dismiss the relevance of a deviation of the
observed radio spectrum from the form $S_{\nu}\propto \nu^{-0.5}$, and of
a likely thermal contribution to the soft X-ray flux around $10^{17}$ Hz,
and try to reproduce all the observed emissions by effects of electrons
alone (e.g. Pohl \cite{pohl}; Mastichiadis \& de~Jager \cite{mast96};
Yoshida \& Yanagita \cite{yoshida}; Aharonian \& Atoyan \cite{aha99}). 
The number of injected and
accelerated protons and all the effects which they can produce in SN~1006
is suggested to be negligibly small in this extreme case. For example, the
proton contribution to the TeV $\gamma$-ray emission should be at least an
order of magnitude lower compared with the previous case. At such low
proton numbers the nonlinear shock modification due to their backreaction
is negligible, and shock acceleration takes place in the test particle
limit.

In order to make clear how different the required set of relevant physical
parameters compared to the case of efficient CR nucleon acceleration is,
we performed a calculation which corresponds to inefficient CR
acceleration. We use a proton injection rate $\eta=10^{-5}$ which yields
an upper limit for the proton acceleration efficiency that is consistent
with their low contribution to the $\gamma$-ray production. This low CR
nucleon production efficiency also implies that the effect of magnetic
field amplification due to CR streaming is small. Therefore we have used a
typical upstream ISM magnetic field strength $B_0=4$~$\mu$G. We note that
such a low value was even assumed to be required in the compressed {\it
downstream} region -- $B_{\perp}=4$~$\mu$G -- to fit the experimental data
(Tanimori et al. \cite{tan01}). All other ISM and SN parameters are the
same as in the previous case of efficient CR acceleration. Since during
the early evolutionary phase the global SNR dynamics is rather insensitive
to the number of accelerated CRs, the observed SNR size and its expansion
rate are equally well reproduced in this case.

In this test particle regime the electron spectrum, produced by a strong
unmodified shock, can be analytically approximated by (Berezhko
\cite{ber96})
\begin{equation} N_\mathrm{e}\propto
p^{-\gamma}\exp[-\kappa(p)/\kappa(p_\mathrm{max}^\mathrm{e})],
\label{eq18} 
\end{equation} 
with $\gamma\approx 2$, where $p_\mathrm{max}^\mathrm{e}$ is the maximum
CR momentum, which is determined by geometrical factors. Any essential
deviation from $\gamma\approx 2$ would be inconsistent with diffusive
shock acceleration which is believed to be the main process of energetic
particle production in SNRs. For a Bohm-type diffusion coefficient,
$\kappa \propto p$, the cutoff region of this spectrum has the same form
$N_\mathrm{e} \propto \exp(-p/p_\mathrm{max}^\mathrm{e})$ as it was
suggested by Mastichiadis \& de Jager (\cite{mast96}) and by Tanimori et
al. (\cite{tan01}).  At the same time, cf.  Eq. (\ref{eq18}), the spectrum
of the electrons accelerated at a strong unmodified shock is essentially
harder compared to what was used $(\gamma=2.2)$ to fit the radio, X-ray
and $\gamma$-ray data (Tanimori et al. \cite{tan01}). It is clear that the
spectrum cf. Eq. (\ref{eq18}), with $\gamma=2$ which has the same number
of GeV electrons to produce the same radio flux, has an order of magnitude
more electrons with $\epsilon_\mathrm{e}=10$~TeV compared with the case
$\gamma=2.2$. Therefore, to fit the radio, X-ray, and $\gamma$-ray data
with this harder electron spectrum, we need a large downstream magnetic
field $B_{\perp} \approx 13$~$\mu$G. Since even for $\eta=10^{-5}$ nucleon
acceleration can produce a strong Alfv\'{e}nic wave field, the upper limit
for the downstream field strength $B_\mathrm{d}$ equals 16~$\mu$G. To
this extent $B_{\perp} \approx 13$~$\mu$G can still be consistent with a
4~$\mu$G ISM magnetic field as used by Tanimori et al.(\cite{tan01}),
although these authors used 4~$\mu$G also downstream.
\begin{figure} 
\centering 
\includegraphics[width=7.5cm]{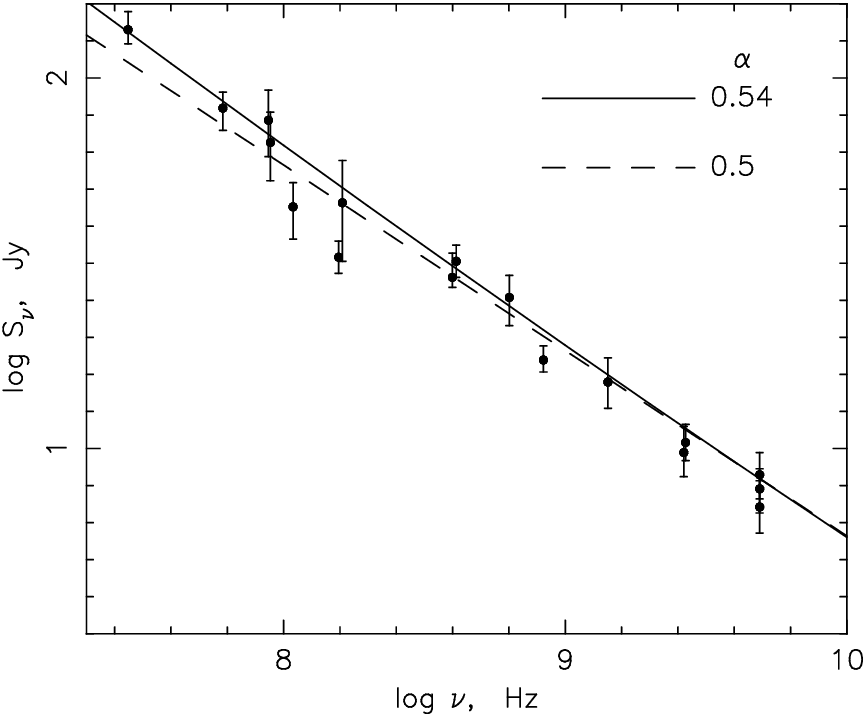}
\caption{Observed total radio flux of SN~1006 (Reynolds \cite{reyn96}) as
a function of frequency with model spectra superimposed. Solid and dashed
curves correspond in this figure to the high and low proton
injection/acceleration
efficiency, respectively.} 
\label{f6} 
\end{figure}
Since the energies $\epsilon_\mathrm{e}
\,\,\raisebox{0.2em}{$>$}\!\!\!\!\! \raisebox{-0.25em}{$\sim$}\,\,
10^{13}$~eV of electrons, whose synchrotron emission in the given magnetic
field corresponds to X-ray energies $\epsilon_{\nu}>5$~keV, should be in
the exponential cutoff region, the required value of the maximum CR
momentum is $p_\mathrm{max}^\mathrm{e}=10^4m_\mathrm{p}c$.  If we assume
the same momentum dependence $\kappa \propto p$, this maximum CR momentum
is consistent with the diffusion coefficient
\begin{equation}
\kappa(p)=5.3\kappa_\mathrm{Bohm}(p), \label{eq19}
\end{equation}
which exceeds the Bohm limit by a factor 5.3 for an upstream field of 4~$\mu$G.

\begin{figure} 
\centering 
\includegraphics[width=7.5cm]{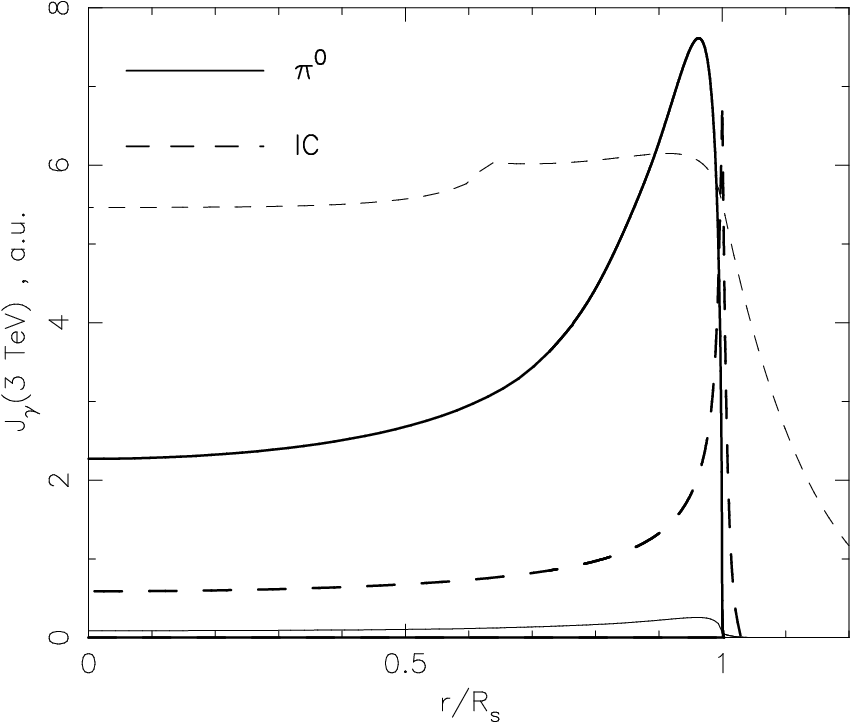}
\caption{Radial dependence of the $\gamma$-ray brightness for the
$\gamma$-ray energy $\epsilon_{\gamma}=3$~TeV. Thick and thin curves
correspond to the high and low proton injection/acceleration efficiency,
respectively.} 
\label{f7} 
\end{figure}

The results for this inefficient proton injection/acceleration case are
also presented in Fig.\ref{f2}--\ref{f7}. One can see from Fig.\ref{f2}
that, due to the lower magnetic field value, one needs about five times
more accelerated electrons compared with the previous case to fit the
radio and X-ray data (see Fig.\ref{f3}).

In the inefficient case the $\gamma$-ray production is dominated by the
electron contribution (Fig.\ref{f4}). Since synchrotron losses are not
important, the predicted IC spectrum is essentially harder for
$\gamma$-ray energies between about $10^9$ and $10^{11}$ eV than in the
efficient proton acceleration model. The differential $\gamma$-ray
CANGAROO spectrum appears in better agreement with the $\pi^0$-decay
emission of the efficient model than with the inefficient model IC
prediction (Fig.\ref{f5}).

In Fig.\ref{f6} the calculated synchrotron fluxes are compared with the
experimental data in the radio range. One can see that the steeper
spectrum $S_{\nu}\propto \nu^{-0.54}$ which corresponds to the efficient
CR acceleration case gives a better fit to the radio data than the
spectrum $S_{\nu}\propto \nu^{-0.5}$ corresponding to the case of
inefficient proton injection, even though the experimental accuracy does
not very clearly distinguish between these to variants. Note that
$\alpha=0.54$ is the average value of the power-law index within the
frequency range shown in Fig.\ref{f6}. In fact, due to the concave shape
of the electron spectrum, the index slightly decreases with increasing
frequency, with $\alpha =0.56$ and $\alpha=0.51$ at the lowest and largest
frequency, respectively.

A similar situation is encountered in the X-ray range. An essentially
steeper electron spectrum in the case of efficient CR accelerations much
better fits the nonthermal X-ray data by lying below the total X-ray flux
near the frequency $\nu = 10^{17}~\mathrm{Hz}$ which corresponds to the
photon energy $\epsilon_{\nu}= 0.4~\mathrm{keV}$ (Fig.\ref{f3}).

At given number of protons and magnetic field strength an electron to
proton ratio $K_\mathrm{ep}=0.04$ is required in the inefficient case, in
order to reproduce the observed emission.

Nucleonic CRs absorb in this case only 
\begin{equation}
E_\mathrm{c}=0.008E_\mathrm{sn}=2.4\times 10^{49}~\mbox{erg}, \label{eq20}
\end{equation}
which is more than an order of magnitude less than the canonically
required $0.1E_\mathrm{sn}$ for the Galactic CRs. The electron CR
component contains $E_\mathrm{c}^\mathrm{e}\approx 10^{48}$~erg. This is
a factor 7 less than what has been obtained by Dyer et al. (\cite{dyer})
corresponding to their lower downstream magnetic field.

It is important to note that the $\gamma$-ray brightness of the remnant 
\begin{equation} J_{\gamma}(\epsilon_{\gamma},r)\propto \int dx
q_{\gamma}(\epsilon_{\gamma},r,x) 
\label{eq21} 
\end{equation} 
has a very different dependence upon the distance from its center $r$ in
the two considered cases. The integration of the $\gamma$-ray emissivity
$q_{\gamma}$ (which is the number of photons emitted per unit volume per second
in the range $\epsilon_{\gamma}$ to $\epsilon_{\gamma} +d\epsilon_{\gamma}$) 
in the above expression is performed along the line of sight
which intersects the visible remnant surface at the projected distance $r$
from its center. In the case of inefficient proton injection TeV
$\gamma$-rays are produced by the electrons with the highest energies of
their spectrum $\epsilon_\mathrm{e}\,\,\raisebox{0.2em}{$>$}\!\!\!\!\!
\raisebox{-0.25em}{$\sim$}\,\, \epsilon_\mathrm{max}^\mathrm{e}$. These
electrons almost uniformly occupy the entire volume between the shock
surface $R_\mathrm{s}$ and the contact discontinuity $R_\mathrm{p}$ (which
separates the ejecta and swept-up matter) with only a slight increase of
their density toward the shock position.  Depending on the turbulence
level near the contact discontinuity they also partially fill the ejecta
volume $r \,\,\raisebox{0.2em}{$<$}\!\!\!\!\!
\raisebox{-0.25em}{$\sim$}\,\, R_\mathrm{p}$.  Since the upstream
distribution of these electrons is characterized by a diffusive length
which is of the order of $0.1R_\mathrm{s}$, the region $r>R_\mathrm{s}$
provides a noticeable contribution to the $\gamma$-ray brightness.
Therefore the TeV $\gamma$-ray brightness $J_{\gamma}(r)$ will have a peak
value in the region $R_\mathrm{p}<r<R_\mathrm{s}$ if particle penetration
through the contact discontinuity is strongly suppressed, and at the
center of the remnant $r=0$ in the opposite case.
%
%
A significantly different radial profile $J_{\gamma}(r)$ is expected in
the case of efficient proton production. TeV $\gamma$-rays are produced in
this case by protons whose energy is considerably lower than the cutoff
energy $\epsilon_\mathrm{max}$. Therefore they are concentrated within the
thin region of thickness $\Delta r\approx R_\mathrm{s}/(3\sigma)$ just
behind the shock. The swept-up gas has a similar distribution. Since the
$\pi^0$-decay $\gamma$-ray luminosity is proportional to the product of
the CR density and the gas number density, the $\gamma$-ray brightness
$J_{\gamma}(r)$ is expected to have a sharp peak close to the shock edge
$r\approx R_\mathrm{s}$ with a considerable decrease at small distances
$r\ll R_\mathrm{s}$. 
The morphological effect of renormalization is to 
limit this emission to two polar caps where the interstellar magnetic 
field is parallel to the shock normal.

The $\gamma$-ray brightness calculated for the two different cases is
presented in Fig.\ref{f7} for the $\gamma$-ray energy
$\epsilon_{\gamma}=3$~TeV. One can see the essentially different radial
dependence $J_{\gamma}(r)$ for the two cases. The $\gamma$-rays of
hadronic origin are expected to be concentrated in the postshock region,
whereas the IC $\gamma$-ray brightness, calculated for quite a large level
of turbulence near the contact discontinuity (see Berezhko \& V\"olk
\cite{bv00} for details), is almost uniformly distributed across the
visible disk of the remnant. As it is seen from Fig.7, the IC high energy
emission of the inefficient case produces a noticeable halo around the
shock -- for SN1006 its size is about 0.1 degrees -- that can in principle
be used to discriminate between the $\gamma$-ray emissions of hadronic and
leptonic origin.

The reported $\gamma$-ray flux was detected from the same outer part of
SN~1006 which shows the radio-emission. As it was already pointed out 
(Aharonian \cite{ah99}; Aharonian \& Atoyan \cite{aha99};  Berezhko et al.
\cite{bkp99}; Berezhko et al. \cite{bkv01}), this can be considered as an
observational argument favoring a strong role for the nuclear CR
component.

\section{Summary and Conclusions}

The nonlinear kinetic model for CR acceleration in SNRs has been applied
to SN~1006 in order to explain its observed properties. We have used
stellar ejecta parameters $M_\mathrm{ej}=1.4M_{\odot}$, $k=7$, distance
$d=1.8$~kpc, and ISM number density $N_\mathrm{H}=0.3$~cm$^{-3}$ from X-ray and
optical imaginary of SN~1006.
 
For these parameters an explosion energy $E_\mathrm{sn}=3\times
10^{51}$~erg is required to fit the observed size $R_\mathrm{s}$ and
expansion speed $V_\mathrm{s}$ which are determined by the ratio
$E_\mathrm{sn}/N_\mathrm{H}$.

The number of accelerated electrons required to fit the radio and X-ray
emission of SN~1006, and correspondingly the role of accelerated protons
in $\gamma$-ray production, depends essentially on the magnetic field
value $B_0$.

It was demonstrated that for low magnetic field $B_0=4$~$\mu$G all the
observed emissions can be dominated by the electron contribution. Protons
are then assumed to be injected into the acceleration much less
efficiently than electrons. For this test particle case the lowest
permitted value of the electron to proton ratio is $K_\mathrm{ep}=0.04$.
It exceeds the canonical value 0.01 observed in situ in the neighborhood
of the Solar System for the Galactic CRs. The maximum energy of
accelerated CRs and their total energy content in this case are only
$\epsilon_\mathrm{\max}\sim 10^{13}$~eV and $E_\mathrm{c}< 3\times
10^{49}$~erg respectively. These numbers are too low for such SNRs to be
considered as the main sources of the nucleonic Galactic CRs.

If CRs in SN~1006 are produced due to the diffusive shock acceleration
process, then even in the case of inefficient proton injection quite a
large but plausible {\it downstream} magnetic field $B_\mathrm{d}\approx
13$~$\mu$G is required to fit the data. It is several times larger than
assumed in a simple estimate by Tanimori et al. (\cite{tan01}), because
the shock produces in this case an electron spectrum $N_\mathrm{e}\propto
\epsilon_\mathrm{e}^{-2}$ which is significantly harder than the spectrum
assumed in that estimate.

The existing SNR data are better approximated if a significantly larger
upstream magnetic field value $B_0=20$~$\mu$G and a physically much more
plausible, efficient nucleon injection rate are assumed. Such an ion
injection rate is estimated from injection theory, and consistent with the
observed radio spectral index. The required magnetic field strength, that
is significantly higher than the rms value $5~\mu$G in the ISM, might be
the result of non-linear amplification near the SN shock by the CR
acceleration process itself.

The results for this case of efficient proton acceleration are then as
follows: 

We find that after adjustment of the predictions of the nonlinear
spherically-symmetric model by a renormalization of the number of
accelerated nuclear CRs to take account of the large area of
quasiperpendicular shock regions in a SNR, good consistency with all
observational data can be achieved, including the reported TeV
$\gamma$-ray flux. The $\pi^0$-decay $\gamma$-ray flux produced by the
nuclear CR component exceeds the flux of IC $\gamma$-rays generated by the
electronic CR component at all energies above about 100 MeV. The theory
did not make use of any knowledge to be derived from $\gamma$-ray
measurements. Therefore the reported TeV flux from SN~1006 supports the
idea that the nuclear CR component is indeed produced in SNRs. The
$\pi^0$-decay $\gamma$-ray flux comes from two polar caps of the remnant.

The maximum energy of accelerated protons
$\epsilon_\mathrm{max}=3\times10^{14}$~eV and their total energy content
$E_\mathrm{c}\approx 3\times 10^{50}$~erg, reproduced in this case, are
consistent with the requirements for the Galactic CR sources. The electron
to proton ratio of $K_\mathrm{ep}= 1.5\times 10^{-3}$, on the other hand,
is lower than the canonical value 0.01.

Comparing the case of efficient proton acceleration with the inefficient
proton acceleration case, we see that the expected $\pi^0$-decay
$\gamma$-ray flux $F_{\gamma}^{\pi}\propto \epsilon_{\gamma}^{-1}$ extends up to
almost 100~TeV, whereas the IC $\gamma$-ray flux reaches less than about
10 TeV. Therefore the detection of $\gamma$-ray emission above $10$~TeV
would imply evidence for a hadronic origin.

We therefore conclude that the analysis of SN 1006 on the basis of overall
SNR dynamics and nonlinear diffusive shock acceleration theory results in
a picture where the nuclear component is strongly accelerated, consistent
with all data for this SNR.

The ratio of accelerated electrons to protons comes out lower by a factor
of about 6 than the canonical ratio of 0.01, generally deduced from
observations of the Galactic CRs. If SN 1006 is typical for the nuclear
Galactic CR source population, then other sources like young Pulsars or
Pulsar Nebulae must also significantly contribute to the Galactic CR
electron population. The Crab Nebula is a point in case.  Since the
majority of Galactic Supernovae is not of the type Ia to which SN 1006
belongs, but rather occurs as a consequence of core collapse events, there
is the alternative possibility that core collapse Supernovae produce a
higher electron to proton ratio.

Phenomenological studies of the $\gamma$-ray emission from SN 1006 on the
basis of the observed synchrotron emission have preferred a dominance of
IC emission from electrons in the $\gamma$-ray part of the emission
spectrum. This scenario can not explain the apparent bipolar morphology
inferred from the existing $\gamma$-ray measurements and yields a much
less convincing approximation to the radio and X-ray synchrotron spectrum.
In addition, as we have shown, the IC emission should reach at best about
10 TeV. Apart from future detailed determinations of the $\gamma$-ray
morphology one therefore needs to precisely measure the $\gamma$-ray flux
at energies $\epsilon_{\gamma}$ between $100$~GeV and $100$~TeV, and
preferably even from 10 MeV upwards.
The detailed spectra and in particular the existence of $\gamma$-rays with
very high energies above $10$~TeV should allow a confirmation, or a
rejection, of CR nucleon production in SN~1006 with an
acceleration efficiency that is consistent with the requirements on the
Galactic CR energy budget.

\begin{acknowledgements}
 This work has been supported in
part by the Russian Foundation for Basic Research (grants 00-02-17728,
99-02-16325). EGB and LTK acknowledge the hospitality of the
Max-Planck-Institut f\"ur Kernphysik, where part of this work was
carried out. 
\end{acknowledgements}

\end{document}